\def\ANON{0} 

\documentclass[runningheads]{llncs}
\usepackage{bbding}
\usepackage{graphicx}
\usepackage{booktabs}
\usepackage{url}
\usepackage{xcolor}
\newcommand\blfootnote[1]{%
  \begingroup
  \renewcommand\thefootnote{}\footnote{#1}%
  \addtocounter{footnote}{-1}%
  \endgroup
}

\begin{document}

\title{The Limit Order Book Recreation Model (LOBRM): An Extended Analysis}
\titlerunning{Exploring the LOB Recreation Model}
%

\ifnum\ANON=1
    \author{Anon}
    \institute{Institute \\ Address}
\else
    \author{Zijian Shi  \Envelope \inst{1}\orcidID{0000-0001-7823-8527} \and
John Cartlidge\inst{1}\orcidID{0000-0002-3143-6355}}
    \authorrunning{Z. Shi and J. Cartlidge}
    %
    \institute{Department of Computer Science, University of Bristol, Bristol BS8 1UB, UK \\
    \email{\{zijian.shi,john.cartlidge\}@bristol.ac.uk}}
    
\fi

\maketitle              
\begin{abstract}
The limit order book (LOB) depicts the  fine-grained demand and supply relationship for financial assets and is widely used in market microstructure studies. Nevertheless, the availability and high cost of LOB data restrict its wider application. The LOB recreation model (LOBRM) was recently proposed to bridge this gap by synthesizing the LOB from trades and quotes (TAQ) data. However, in the original LOBRM study, there were two limitations: (1) experiments were conducted on a relatively small dataset containing only one day of LOB data; and (2) the training and testing were performed in a non-chronological fashion, which essentially re-frames the task as interpolation and  potentially introduces lookahead bias. In this study, we extend the research on LOBRM and further validate its use in real-world application scenarios. We first advance the workflow of LOBRM by (1) adding a time-weighted z-score standardization for the LOB and (2) substituting the ordinary differential equation kernel with an exponential decay kernel to lower computation complexity. Experiments are conducted on the extended LOBSTER dataset in a chronological fashion, as it would be used in a real-world application. We find that (1) LOBRM with decay kernel is superior to traditional non-linear models, and module ensembling is effective; (2) prediction accuracy is negatively related to the volatility of order volumes resting in the LOB; (3) the proposed sparse encoding method for TAQ exhibits good generalization ability and can facilitate manifold tasks; and (4) the influence of stochastic drift on prediction accuracy can be alleviated by increasing historical samples.

\keywords{Limit order book  \and Time series prediction \and Financial machine learning.}
\end{abstract}

\section{Introduction}\label{sec1}\blfootnote{This is a preprint manuscript accepted for publication at ECML-PKDD 2021.}

\noindent 
The majority of financial exchange venues utilise a continuous double auction (CDA) mechanism \cite{friedman1993double} for matching orders. Under CDA formation, both {\em ask} orders (orders to sell a given quantity at a given price) and {\em bid} orders (orders to buy a given quantity at a given price) arrive at the venue continuously, with no minimum time interval limit. When a new order arrives, if it does not immediately execute, it will enter the limit order book (LOB); which contains a list of current bids and a list of current asks, both sorted by price-time priority. Therefore, the LOB contains valuable information on the instantaneous demand and supply for a particular financial asset (e.g., a stock, a commodity, a derivative, etc.). For this reason, LOB data has been used for many and various studies, including exploration of the price formation mechanism \cite{parlour1998price}, market anomaly detection \cite{ye2019extracting}, and testing of  trading algorithms \cite{abergel2020algorithmic}.

However, there remain some obstacles for the wider application of LOB data. Firstly, LOB data subscription fees are usually high, sometimes amounting to tens of thousands of dollars per annum.\footnote{http://www.nasdaqtrader.com/Trader.aspx?id=DPUSdata} This might be a trivial sum for an institutional subscriber, however for individual investors and researchers this significant expense can hold them back. Further, LOB data is entirely unavailable in venues that deliberately do not make order information public, for instance some e-commercial markets and {\em dark pools} (e.g., see \cite{MPC-CarSmaTal-19,EPRINT:CarSmaTal20}). This challenge attracts researchers to consider the possibility of recreating the LOB from a more easily available source, such as {\em trades and quotes} (TAQ) data. TAQ data contains the {\em top price level} information of a LOB (the lowest-priced ask and highest-priced bid), together with a history of transactions. It is published to the public for free in most venues. Blanchet et al. \cite{blanchet2017unraveling} have previously demonstrated that it is possible to predict daily average order volumes resting at different price levels of the LOB, using only TAQ data for parameter estimation. More recently, from a deep learning perspective, the LOB recreation model (LOBRM) was proposed to formalize the task as a time series prediction problem, and an ensembled recurrent neural network (RNN) model was successfully used to predict order volumes in a high frequency manner for the first time \cite{shi2021lob}. Nevertheless, there exist two key restrictions in the LOBRM study: (1) The original LOBRM study was conducted in an interpolation style on only one day's length of LOB data, for two stocks. For the model to be applied in a real world application scenario, such as online prediction of market price movements, LOBRM performance requires evaluation on an extended multi-day dataset, with chronological training and testing such that there is no possibility of lookahead bias; (2) The ordinary differential equation (ODE) kernel used in the original LOBRM model has high computation complexity and is therefore inefficient for more realistic application scenarios when large amounts (weeks or months) of training data is used.

\paragraph{\bf Contributions:}
\begin{enumerate}
\item We advance the workflow and structure of the LOBRM model, such that: (i) a time-weighted z-score standardization for LOB features is used to enhance the model's generalization ability; and (ii) the original ODE kernel is substituted for an exponential decay kernel to enable faster inference of latent states, greater runtime efficiency, and a reduction in overfitting. 

\item We use chronological training and testing to conduct experiments on an extended LOBSTER dataset that is an order of magnitude larger than the original dataset. We find that: (i) LOBRM with continuous decay kernel is superior in modelling the irregularly sampled LOB; and (ii) the module ensembling of LOBRM is effective.
\item We draw new empirical findings that further enrich the current literature: (i) the proposed sparse encoding method for TAQ data has good generalization ability and can facilitate manifold tasks including LOB prediction and price trend prediction; (ii) prediction accuracy of the LOBRM is negatively related to volume volatility at unseen price levels; and (iii) the influence of stochastic drift on model performance can be alleviated by increasing the amount of historical training samples.
\end{enumerate}

\section{Background and Related Work}\label{sec2}
\noindent
\subsection{The Limit Order Book (LOB)}
\begin{figure}[t]
  \centering
  \includegraphics[width=0.85\linewidth]{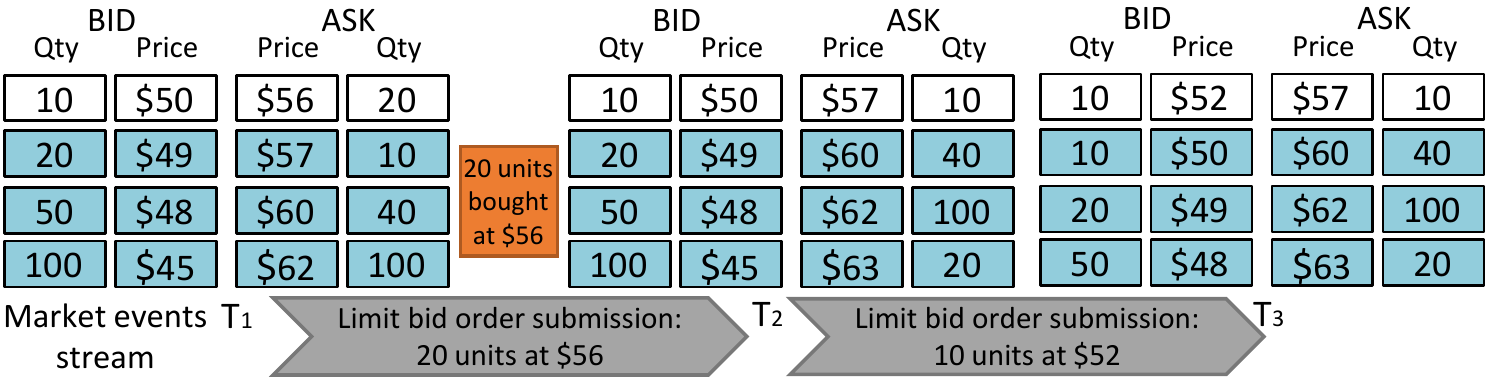}
  \caption{A LOB of four price levels evolving with time. White and blue boxes indicate the top level and deeper levels of the LOB. Grey boxes indicate market events stream. Orange box indicates trade records. White and orange boxes together form TAQ.}
  \label{fig:LOB}
\end{figure}

In a CDA market, bids and asks with specified price and quantity (or {\em volume}) are submitted, cancelled, and transacted continuously. The 
LOB contains an ask side and a bid side, with ask orders arranged in price ascending order and bid orders arranged in price descending order. Ask orders with the lowest price ({\em best ask}) and bid orders with the highest price ({\em best bid}) form the top level of a LOB, and their respective prices are called {\em quotes}. If a newly submitted ask (or bid) price is not higher (or lower) than the best bid (or ask), a trade happens. TAQ data contains all historical quotes and trades in the venue. That is, LOB data contains strictly more information than TAQ data. Fig.~\ref{fig:LOB} provides a visual illustration of a LOB and the relationship between the LOB and the TAQ data. 

Traditional statistical models of the LOB assume that LOB evolution follows the rules of a Markovian system, with market events (order submission, cancellation, and transaction) following stochastic point processes, such as a Hawkes process or a Poisson process \cite{abergel2015long,cont2013price}. This formulation generalizes a LOB market of high complexity to a dynamic system controlled by a few parameters, where probabilistic theorems like the law of large numbers \cite{horst2017weak} and stationary equilibrium in a Markovian system \cite{blanchet2017unraveling} can be utilized to draw long-term empirical conclusions. However, while statistical modelling can capture long term behaviour patterns of the LOB, these models cannot consistently perform well in a high frequency domain. 
In recent years, there has been an emergence of research using deep learning approaches to model and exploit the LOB. Sirignano et al. \cite{sirignano2019universal} performed a significant study on a comprehensive pool of 500 stocks. They revealed that features learned by a Long Short-Term Memory (LSTM) network can be utilized to predict next mid-price movement direction $\{$\textit{up, down}$\}$ with accuracy range $[0.65,0.76]$ across all 500 stocks. Their study also demonstrated that deep learning models suffer less from problems such as stochastic drift that exist in statistical models of the LOB. Other deep learning studies of the LOB include extracting high frequency indicators \cite{passalis2020temporal,tsantekidis2020using}, predicting future stock price movement \cite{makinen2019forecasting,zhang2019deeplob}, and training reinforcement trading agents \cite{kumar2020deep,nevmyvaka2006reinforcement}.

\subsection{Generating Synthetic LOB Data}
\noindent
Synthetic LOB data, generated by models that learn from the real LOB or imitate the {\em stylized facts} of a CDA market, has been used as an alternative when real LOB data is unavailable. The advantages of using synthetic LOB data lie in its low cost and infinite availability. It has been widely adopted to backtest trading algorithms, explore market dynamics, and facilitate teaching activities. 

Synthetic LOB data can be generated using three mainstream methodologies. 
\textbf{Agent-based models} have been well studied and are the most popular approach for generating a synthetic LOB.
By configuring agents that trade using common strategies, such as market makers, momentum traders, and mean reversion traders, the synthetic LOB can closely approximate the stylised facts of a real LOB \cite{mcgroarty2019high}. Stock market simulators have a long history, from the Santa Fe artificial stock exchange \cite{arthur1996asset} to recent multi-agent exchange environments \cite{belcak2020fast}. 
\textbf{Generative models} attempt to learn regularities embedded in market event streams or the LOB directly. One representative research by Li et al. \cite{li2020generating} utilized generative adversarial networks to learn and replicate the historical dependency among orders. The synthesised order stream and resulting LOB were found to  closely resemble the real market data. 

We consider the aforementioned two approaches as {\em unsupervised}, since no real LOB data is used to verify the authenticity of the generated data. Model quality can only be verified by testing whether certain {\em stylized facts} exist in the synthetic data. 
In contrast, \textbf{supervised models} use real LOB data as ground truth. As indicated in \cite{blanchet2017unraveling}, TAQ data is informative of LOB volumes for small-tick stocks. By modelling the tail probability of price change per trade in a Markovian LOB, daily average order volumes at unseen price levels can be estimated by the steady-state distribution of the infinite server queue. Shi et al. \cite{shi2021lob} further formulate the task of generating a synthetic LOB as a time series prediction problem using a continuous RNN. The proposed model (LOBRM) is able to predict LOB order volumes using a defined length of TAQ data as input. As long as a historical TAQ trajectory is available, the model is able to produce a historical replay of the LOB based on the knowledge it learned from supervised training. This paper concentrates on a further exploration of the LOBRM model presented in \cite{shi2021lob}.

\section{Model Formulation}\label{sec3}
\subsection{Motivation}\label{sec:motivation}
\noindent
The LOBRM model represents the first attempt to synthesize LOB data from a supervised deep learning perspective \cite{shi2021lob}. LOBRM is essentially an ensemble of RNNs that take TAQ data as input and produce LOB volume predictions as output. However, in the original study, there were three restrictions present: 
    (1) 
    Experiments were performed using a relatively small LOB dataset consisting of only one day's LOB data for two small-tick stocks. To verify its generalization ability, LOBRM requires testing on multi-day LOB data for a variety of stocks.
    (2) Experiments adopted a non-chronological approach to the formation of time series samples, such that samples were shuffled before splitting into training and testing sets. We intend to test model performance using a strictly chronological approach to ensure that LOBRM is applicable to real world online scenarios, with no possibility of introducing lookahead bias. Specifically, we use the first three days' data for training, the fourth day's data for validation, and the fifth day's data for testing. Time series samples are not shuffled, thus ensuring that chronological ordering is preserved.
    (3) The core module of the original LOBRM made use of an ODE-RNN, a RNN variant with ODE kernels to derive fine-grained time-continuous latent states \cite{rubanova2019latent}. As the computational complexity of ODE-RNN is $n$ times that of a vanilla RNN -- where $n$ is the granularity of the latent state -- it is not efficient for training data of large size. Therefore, we substitute the ODE-RNN for an RNN-decay module \cite{che2018recurrent}, which has been shown to be a more time-efficient model for irregularly sampled time series \cite{lechner2020learning}. 

\subsection{Problem Description}
\noindent
To simplify the problem of recreating the LOB, we make the same assumptions proposed in \cite{shi2021lob}: (1) Following common practice (e.g., \cite{blanchet2017unraveling,horst2017weak}), the bid and ask sides of the LOB are modelled separately; (2) We only consider instantaneous LOB data at the time of each trade event, and ignore the LOB for all other events, such as order submission and cancellation; (3) We only consider the top five price levels of the LOB; (4) We assume that the price interval between each price level is exactly one {\em tick} -- the smallest increment permitted in quoting or trading a security at a particular exchange venue -- which is supported by empirical evidence that orders in the LOB for small-tick stocks tend to be densely distributed around the top price levels \cite{blanchet2017unraveling}. Following these assumptions, the price at different LOB levels can be directly deduced from the known quote price at target time. Therefore, the LOB recreation task resolves to the simpler problem of only predicting order volumes resting at each price level.

For generalization, we denote trades and quotes streams as $\left \{ TD_{i} \right \}_{i\in \textsl{n}}$ and $\left \{ QT_{i} \right \}_{i\in \textsl{n}}$ respectively, 
and trajectories of time points for TAQ records as $\left \{ T_{i} \right \}_{i\in \textsl{n}}$ indexed by $\textsl{n}=\left \{ 1,\ldots,N \right \}$, where $N$ equals the number of time points in TAQ. 
The LOB sampled at $\left \{ T_{i} \right \}_{i\in \textsl{n}}$ are denoted as $\left \{ LOB_{i} \right \}_{i\in \textsl{n}}$. For each record at time $T_{i}$, $QT_{i}=\left( p_{i}^{a(1)}, v_{i}^{a(1)}, p_{i}^{b(1)}, v_{i}^{b(1)}\right)$, where $p_{i}^{a(1)}, v_{i}^{a(1)}, p_{i}^{b(1)}, v_{i}^{b(1)}$ denote best ask price, order volume at best ask, best bid price, and order volume at best bid, respectively. 
$TD_{i}=\left(p_{i}^{td}, v_{i}^{td}, d_{i}^{td}\right)$, where $p_{i}^{td}, v_{i}^{td}, d_{i}^{td}$ denote price, volume, and direction of the trade, with $+1$ and $-1$ indicating orders being sold or bought.
$LOB_{i}=(p_{i}^{a(l)}, v_{i}^{a(l)}, p_{i}^{b(l)}, v_{i}^{b(l)})$ depicts the price and volume information at all price levels, with $l\in(1,...,L)$, here $L=5$. 
From the aforementioned model assumptions, we have $p_{i}^{a(l)}=p_{i}^{a(1)}+(l-1)\tau$ and $p_{i}^{b(l)}=p_{i}^{b(1)}-(l-1)\tau$, where $\tau$ is the minimum tick size (1 cent in the US market). For a single sample, the model predicts $(v_{I}^{a(2)},...,v_{I}^{a(L)})$ and $(v_{I}^{b(2)},...,v_{I}^{b(L)})$ conditioned on the observations of $\left \{ QT_{i} \right \}_{{I-S}:I}$ and $\left \{ TD_{i} \right \}_{{I-S}:I}$, with $S$ being the time series sample size, i.e., the maximum number of time steps that the model looks back in TAQ data history. 

\subsection{Formalized Workflow of LOBRM} \label{subsec:model}
\noindent
\subsubsection{LOB Data Standardization}
As we intend to apply the LOBRM model on LOBs of five days' length for different financial assets, data standardization is necessary for the model's understanding of data of various numerical scales. We perform time-weighted z-score standardization on all LOB volumes, based on the fact that the LOB is a continuous dynamic system with uneven time intervals between updates. 
We use $\left \{v_{n}^{(l)}\right \}_{0:N}$ and $\left \{T_{n}\right \}_{0:N}$ to indicate volume trajectory on price level $l$, and affiliated timestamps $N$ as the total number of LOB updates for training. Time-weighted mean and standard deviation are calculated as:
\begin{equation}
    \mathrm{Mean}^{(l)} = \begin{matrix}\sum_{i=0}^{N-1}{\left ( T_{i+1}-T_{i} \right ) v_{i}^{(l)}} \end{matrix}/\left (T_{N}-T_{0}\right )\\
\end{equation}
\begin{equation}
    \mathrm{Std}^{(l)} = \left ( \begin{matrix}\sum_{i=0}^{N-1}{\left ( {T_{i+1}-T_{i}} \right )} \left ( v_{i}^{(l)}-\mathrm{Mean}^{(l)} \right )^{2} \end{matrix}/\left (T_{N}-T_{0}\right )  \right )^{\frac{1}{2}} \\
\end{equation}

From empirical observation, we witness that the volume statistics on deeper price levels $\left \{v_{n}^{(2-5)}\right \}_{0:N}$ have similar patterns, while those statistics deviate from volume statistics on the top price level $\left \{v_{n}^{(1)}\right \}_{0:N}$. In particular, the top level tends to have much lower mean and much higher standard deviation (e.g., see later  Table~\ref{tab:dataset statistics}). Thus, we treat `top' and `deeper' levels as two separate sets to standardize. As trade volumes $\left \{v_{n}^{tq}\right \}_{0:N}$ are discrete events and do not persist in time, we use a normal z-score standardization for trade data. To avoid lookahead bias, statistics are calculated without considering test data. Finally, using these statistics, LOBs for training, validation, and testing are standardized. 

\subsubsection{Sparse Encoding for TAQ}
TAQ data contains multi-modal information, including order type ({\em bid} or {\em ask}), price, and volume. While under the formulation of LOBRM, only order volumes at derived price levels (i.e., {\em deeper} levels 2-5) are predicted. We use a one-hot positional encoding, such that only volume information is encoded explicitly; while price is indicated by the position of non-zero elements in the one-hot vector.

Take the encoding of an ask quote as an example. Conditioned on current best ask price $p_{I}^{a(1)}$ and best bid price $p_{I}^{b(1)}$ at $T_{I}$, we represent the ask quote record $(p_{I-s}^{a(1)},v_{I-s}^{a(1)})$ and bid quote record $(p_{I-s}^{b(1)},v_{I-s}^{b(1)})$ at $T_{I-s}$, $s \in \left \{0,\ldots,S\right \}$ as:
\begin{equation}
    \left\{
    \begin{array}{cl}
        O_{2k-1}^{aq}, \mathrm{where}\,o_{k+sp_{s}^{a}}=v_{I-s}^{a(1)} \\
        O_{2k-1}^{bq}, \mathrm{where}\,o_{k+sp_{s}^{b}}=v_{I-s}^{b(1)}
    \end{array} \right.
    \label{eqt_5}
\end{equation}
where $k\in R$, $sp_{s}^{a}=(p_{I-s}^{a(1)}-p_{I}^{a(1)})/\tau$ and $sp_{s}^{b}=(p_{I-s}^{b(1)}-p_{I}^{b(1)})/\tau$. $O_{2k-1}$ is a one-hot vector with dimension $1\times (2k-1)$; and $o_{sp}$ denotes the ${sp}$-th element of the vector. 
The value of $k$ is chosen to cover more than 90\% of past quote price fluctuations, relative to the current quote price. Here $k=8$, which means historical quotes with relative price $[-7,+7]$ ticks are encoded into feature vectors.
Then, a trade record $(p_{I-s}^{td}, v_{I-s}^{td},  d_{I-s}^{td})$, is represented as:
\begin{equation}
    \left\{
    \begin{array}{cl}
        O_{2k-1}^{td}, \mathrm{where}\,o_{p_{I-s}^{td}-p_{I}^{a(1)}}=v_{I-s}^{td} & \mathrm{if} \,d_{I-s}^{td}<0 \\
        O_{2k-1}^{td}, \mathrm{where}\,o_{p_{I-s}^{td}-p_{I}^{b(1)}}=v_{I-s}^{td} & \mathrm{if} \,d_{I-s}^{td}>0 \\
    \end{array} \right.
    \label{eqt_7}
\end{equation}

Finally, those three features are concatenated into $(O^{aq},O^{bq},O^{td})$ and are used as input. It can also be a concatenation of four features, with ask and bid trade represented separately. Later, in experiment section~\ref{sec:encoding}, we show this sparse encoding method can achieve enhanced robustness in LOB volume prediction and price trend prediction.

\subsubsection{Market Event Simulator Module (ES)}
The ES module models the overall net order arrivals as inhomogeneous poisson processes, and predicts LOB volumes from a dynamic perspective. If an RNN structure is used to iteratively receive encoded LOB features at every timestep, its latent state can be deemed as reflective of market microstructure condition over a short historical time window. A multi-layer perceptron (MLP) layer can then be used to decode latent states directly into vectors representing net order arrival rates at each price level.

The original LOBRM model uses ODE-RNN, a continuous RNN variant that learns fine-grained latent state between discrete inputs in a data-driven manner, to model the continuous evolution of market conditions. In ODE-RNN, a hidden state $h(t)$ is defined as a solution to an ODE initial value problem. The latent state between two inputs can then be derived using an ODE solver as:
\begin{equation}
    \begin{array}{cl}
    \frac{dh(t)}{dt}=f_{\theta }(h(t),t) \; \mathrm{where} \; h(t_{0})=h_{0}
    \end{array}
    \label{eqt_1}
\end{equation}
\begin{equation}
    \begin{array}{cl}
        {h_{i}}'=\mathrm{ODEsolver}(f_{\theta}, h_{i-1}, (t_{i-1}, t_{i}))
    \end{array}
    \label{eqt_2}
\end{equation}
in which function $f_{\theta }$ is a separate neural network parameterized by $\theta$.

Even though ODE-RNN contributes most to prediction accuracy, it is of high computation complexity (see Section~\ref{sec:motivation}). 
Therefore, for an efficient use of LOBRM, especially when trained on large amounts of data, or used for online prediction, the ODE-RNN is unsuitable. Also, we find in chronological experiments that the ODE-RNN tends to cause overfitting due to the fully flexible latent states.

Faced with these challenges, we propose the use of a pre-defined exponential decay kernel \cite{che2018recurrent,lechner2020learning}, instead of the ODE kernel in the ES module. The inference of latent states between discrete inputs is denoted as: 
\begin{equation}
    {h_{i}}'= h_{i-1}*exp\left ( -f_{\theta}\left ( h_{i-1} \right ) * \Phi \left ( t_{i-1},t_{i} \right ) \right )
\end{equation}
\begin{equation}
    h_{i} = \mathrm{GRUunit}\left ({h_{i}}',x_{i} \right )
\end{equation}
where $f_{\theta}$ is a separate neural network parameterized by $\theta$, and $\Phi$ is a smoothing function for time intervals to avoid gradient diminishing. A GRU unit \cite{cho2014learning} is then used for instant updating of the latent state at input timesteps. The advantages of employing an exponential decay kernel are threefold: (1) It allows for efficient inference of latent states; (2) It is a continuous RNN that imitates continuity of market evolution and includes temporal information within the model structure itself, sharing the advantages of continuous RNN in modelling irregularly sampled time series; (3) It is less likely to cause overfitting as the kernel form is predefined, whereas the latent state evolution in ODE-RNN is fully flexible.

We derive the vector of net order arrival rates $\Lambda_{I-s}=[\lambda_{I-s}^{a(2)},...,\lambda_{I-s}^{a(L)} ]$ at time $T_{I-s}$ directly from the latent state $h_{I-s}$, using an MLP layer as:
\begin{equation}
    \Lambda_{I-s}=\mathrm{MLP} \left (h_{I-s} \right )
\end{equation}
After acquiring the trajectory of $\Lambda$ at all trade times over the defined length of time steps, we calculate the accumulated order volumes between $[T_{I-S},T_{I}]$ as:
\begin{equation}
    \begin{matrix} \sum_{i=I-S}^{I}\Lambda_i\times \Phi \left (T_{i+1}-T_{i} \right ) \end{matrix}
\end{equation}

\subsubsection{History Compiler Module (HC)}
The HC module predicts LOB volumes from a historical perspective. It concentrates on historical quotes that are most relevant to current LOB volumes at deeper price levels. More precisely, for an ask side model the volumes to be predicted at target time are of prices $\{p_{I}^{a(1)}+\tau, \ldots, p_{I}^{a(1)}+(L-1)\tau\}$. Only historical ask quotes with price within this range are used as inputs into the HC module. 
As one-hot encoded ask quotes fall within the price range of $\{p_{I}^{a(1)}-k\tau, \ldots, p_{I}^{a(1)}+k\tau\}$, vectors need to be trimmed to remove verbose information and retain the most relevant information. Formally, we represent a trimmed ask quote record $(p_{I-s}^{a(1)},v_{I-s}^{a(1)})$ as:
\begin{equation}
    \left\{
    \begin{array}{cl}
    O_{L-1}, \mathrm{where}\,o_{sp_{s}^{a}}=v_{I-s}^{a(1)} & \mathrm{if} \,sp_{s}^{a}\in[1,L-1] \\ Z_{L-1} & \mathrm{otherwise}
    \end{array} \right.
    \label{eqt_3}
\end{equation}
where $O_{L-1}$ is a one-hot vector of dimension $1\times (L-1)$; $o_{sp_{s}^{a}}$ denotes the  
${sp_{s}^{a}}$-th element of the vector; and  $Z_{L-1}$ denotes a zero vector of the same dimension. The intention of the HC module is to look back in history to check how many orders were resting at price levels we are interested in at target time. Thus, we manually trim input feature vectors to leave out verbose information. A discrete GRU unit is used to compile trimmed features and generate volume predictions which are used as supplements to ES predictions. 

\subsubsection{Weighting Scheme Module (WS)}
This module is designed to combine the predictions from the ES and HC modules into a final prediction.
We follow the intuition that if the HC prediction for a particular price level is reliable from a historical perspective, a higher weight will be allocated to it, and {\em vice versa}. If quote history for a target time price level is both abundant and recent, we weight the information  provided on current LOB volumes as more reliable. The abundancy and timing of historical quotes are denoted by the masking sequence of HC inputs. Formally, we represent the mask of an ask side HC input as:
\begin{equation}
    \left\{
    \begin{array}{cl}
    O_{L-1}, \mathrm{where}\,o_{sp_{s}^{a}}=1 & \mathrm{if} \,sp_{s}^{a}\in[1,L-1] \\ Z_{L-1} &  \mathrm{otherwise}
    \end{array} \right.
    \label{eqt_4}
\end{equation}
A GRU unit is used to receive the whole masking sequence to generate a weighting vector of size $1\times\left ( L-1\right )$ to combine predictions from ES and HC modules.

\section{Experiment and Empirical Analysis}\label{sec4}
In this section, experiments are conducted on the extended LOBSTER dataset. These data were kindly provided by \url{lobsterdata.com} for academic research. The stocks and time periods were not selected by the authors. The dataset contains LOB data of five days' length for three small-tick stocks (Microsoft, symbol {\tt MSFT}; Intel, {\tt INTC}; and JPMorgan, {\tt JPM}). To ensure that there is no selection bias or ``cherry-picking'' of data, all three available small-tick stocks were used in this study.  The extended dataset is  approximately ten times the size of the dataset used in \cite{shi2021lob} and is a strict superset, therefore enabling easier results comparison.  

\paragraph{Model specification:} The {\bf ES module} consists of a GRU with 64 units, a two layered MLP with 32 units and ReLU activation for deriving parameters of the decay kernel, and a two layered MLP decoder with 64 units and Tanh activation. The {\bf HC module} consists of a GRU  with 64 units, and a two layered MLP decoder with 64 units and LeakyReLU activation. The latent state size is set at 32 in ES and HC modules. The {\bf WS module} consists of a GRU with 16 units, and a one-layered MLP decoder with 16 units and Sigmoid activation. The latent state size is set at 16 in the WS module. L1 loss is used as the loss function; and all models are trained for 150 iterations, with a learning rate of 2e-4. Model parameters are chosen based on the lowest loss on the validation set.

\subsection{Data Preprocessing}
We clean the dataset by retaining only LOB updates at trade times, and removing LOB data during the first half-hour after market open and the last half-hour before market close as these periods tend to be volatile.
To alleviate the effect of outliers, we divide all volume numbers by 100 and winsorize the data by the range $[0.005,0.995]$. Then, the standardization method proposed in Section~\ref{subsec:model} is applied to the cleansed LOB data. Data statistics are illustrated in Table~\ref{tab:dataset statistics}.

\begin{table}[t]
\footnotesize
\centering
\caption{Volume statistics before standardization, showing time-weighted mean volume and standard deviation on the top level and deep levels (levels 2 to 5).}\vskip -0.5em
\label{tab:dataset statistics}
\setlength\tabcolsep{1.5mm}
\begin{tabular}{l cc cc cc cc}
\toprule
        & \multicolumn{2}{c}{MSFT}  
        & \multicolumn{2}{c}{INTC}
        & \multicolumn{2}{c}{JPM}\\
        \rule{0pt}{2ex}  
        & \multicolumn{1}{c}{Bid} & \multicolumn{1}{c}{Ask} 
        & \multicolumn{1}{c}{Bid} & \multicolumn{1}{c}{Ask} 
        & \multicolumn{1}{c}{Bid} & \multicolumn{1}{c}{Ask}\\
\midrule
Top     & \multicolumn{2}{c}{119.7/89.1} & \multicolumn{2}{c}{127.1/98.3} &  \multicolumn{2}{c}{32.7/38.5}\\
Deep   &  189.6/51.7   &  192.7/53.4  &  185.6/62.3  &  178.9/51.4  &  61.3/23.9  &  63.4/40.5 \\
\bottomrule
\end{tabular}
\end{table}

We use the first three days' data for training, the fourth day's data for validation, and the fifth day's data for testing. This is a significantly different approach to that taken in \cite{shi2021lob}, in which the task was essentially interpolation, as all time series samples were shuffled before splitting into training and testing sets (i.e., training and testing sets were not ordered chronologically). We extract TAQ data and labels directly from the standardized LOB. These data are then converted to time series samples using a rolling window of size $S$, such that the first sample consists of TAQ histories at timesteps $1$ to $S$ and is labeled by LOB volume at deep price levels at time step $S$. The second sample consists of TAQ history at timesteps $2$ to $S+1$ and is labeled using LOB at time step $S+1$, {\em et cetera}.
We set parameters $S=100$ and $k=8$, as in the original LOBRM model \cite{shi2021lob}. 

\subsection{Model Comparison}
\noindent
In this section, we illustrate training results on generating synthetic LOB using mainstream regression and machine learning methods: (1) Support Vector Machine Regression with linear kernel (LSVR); (2) Ridge Regression (RR); (3) Single Layer Feedforward Network (SLFN); and (4) XGBoost Regression (XGBR). We then evaluate the performance of LOBRM with either discrete RNN or Continuous RNN module: (1) GRU; (2) GRU-T, with time concatenated input; (3) Decay; and (4) Decay-T, with time concatenated input.
Two criteria are used: (1) L1 loss on test set. As all labels are standardized into z-score, the loss indicates the multiple of standard deviations between prediction and ground truth; (2) R-squared, calculated on the test data using the method presented by Blanchet et al. \cite{blanchet2017unraveling}, to enable us to perform a strict comparison with the existing literature.

\begin{table}[tb]
\small
\centering
\caption{Model comparison. Criteria shown in format: test loss/R-squared; all numbers are in 1e-1. The lowest test loss and highest R-squared in each set are underlined.}\vskip -0.5em
\setlength\tabcolsep{0.8mm}
\begin{tabular}{l cc cc cc}
\toprule
        Model & \multicolumn{2}{c}{MSFT}  
        & \multicolumn{2}{c}{INTC}
        & \multicolumn{2}{c}{JPM}\\
        \rule{0pt}{2ex}  
        & \multicolumn{1}{c}{Bid} & \multicolumn{1}{c}{Ask} 
        & \multicolumn{1}{c}{Bid} & \multicolumn{1}{c}{Ask} 
        & \multicolumn{1}{c}{Bid} & \multicolumn{1}{c}{Ask}\\
\midrule
LSVR     &  10.00/0.29  &  9.74/0.36    &  8.72/0.21  &  10.30/0.14 &   6.91/0.54   &   3.96/0.51 \\
RR       &  7.90/0.55   &  7.70/0.67    &  6.18/0.53  &  7.00/0.40  &   6.86/0.65   &   4.10/0.46 \\
SLFN     &  7.06/0.68   &  6.88/0.88    &  5.55/0.86  &  5.88/0.53  &   6.61/0.71   &   3.69/0.62 \\
XGBR     &  6.50/0.43   &  6.91/0.80    &  \underline{5.21}/0.82  &  5.93/0.83  &   6.81/0.76    &   3.73/0.69 \\
LOBRM (GRU)    &  6.44/1.24   &  6.44/1.55    &  5.34/1.05  &  5.70/0.53  &   6.27/1.17  &   3.30/1.25  \\
LOBRM (GRU-T)  &  6.45/1.38   &  6.45/1.56    &  5.36/1.06  &  5.65/\underline{0.84}  &   \underline{6.13}/1.48  &   3.34/\underline{1.73}  \\
LOBRM (Decay)  &  6.54/1.30   &  6.47/1.49    &  5.33/\underline{1.11}  &  \underline{5.52}/0.73  &   6.35/1.20  &   \underline{3.29}/1.37  \\
LOBRM (Decay-T)&  \underline{6.28}/\underline{1.58}   &  \underline{6.15}/\underline{1.77}    &  5.33/1.07  &  5.64/0.82  &   6.18/\underline{1.54}  &   3.32/1.65  \\
\bottomrule
\end{tabular}
\label{tab:continuous}
\end{table}

\begin{figure}[t]
\centering
\begin{minipage}[t]{0.45\textwidth}
\centering
\includegraphics[width=0.99\textwidth]{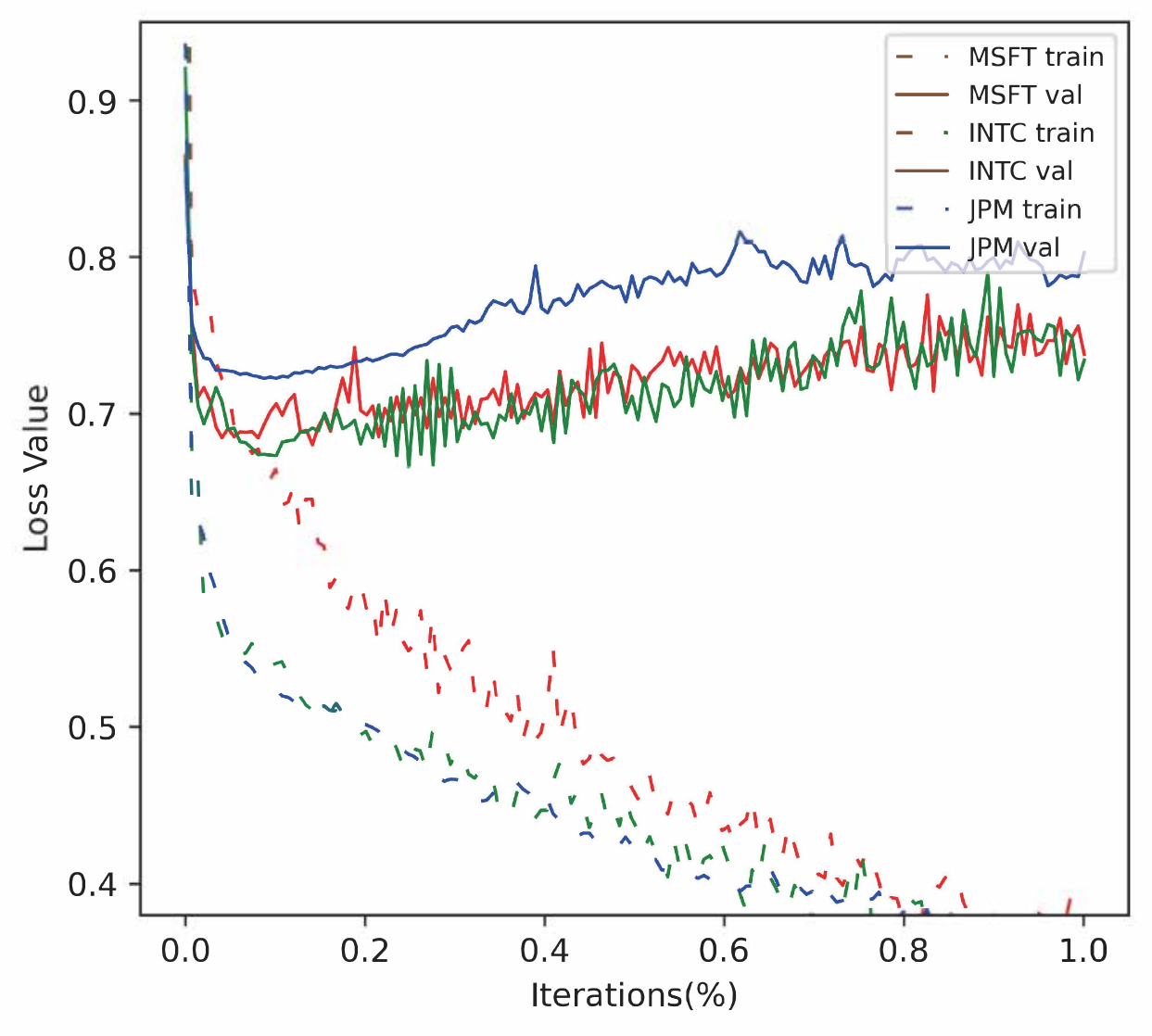}
\vskip -0.5em
\caption{Training (dash) and validation (line) loss curves for bid-side models.}
\label{fig:loss}
\end{minipage}
\hspace{1em}
\begin{minipage}[t]{0.45\textwidth}
\centering
\includegraphics[width=0.99\textwidth]{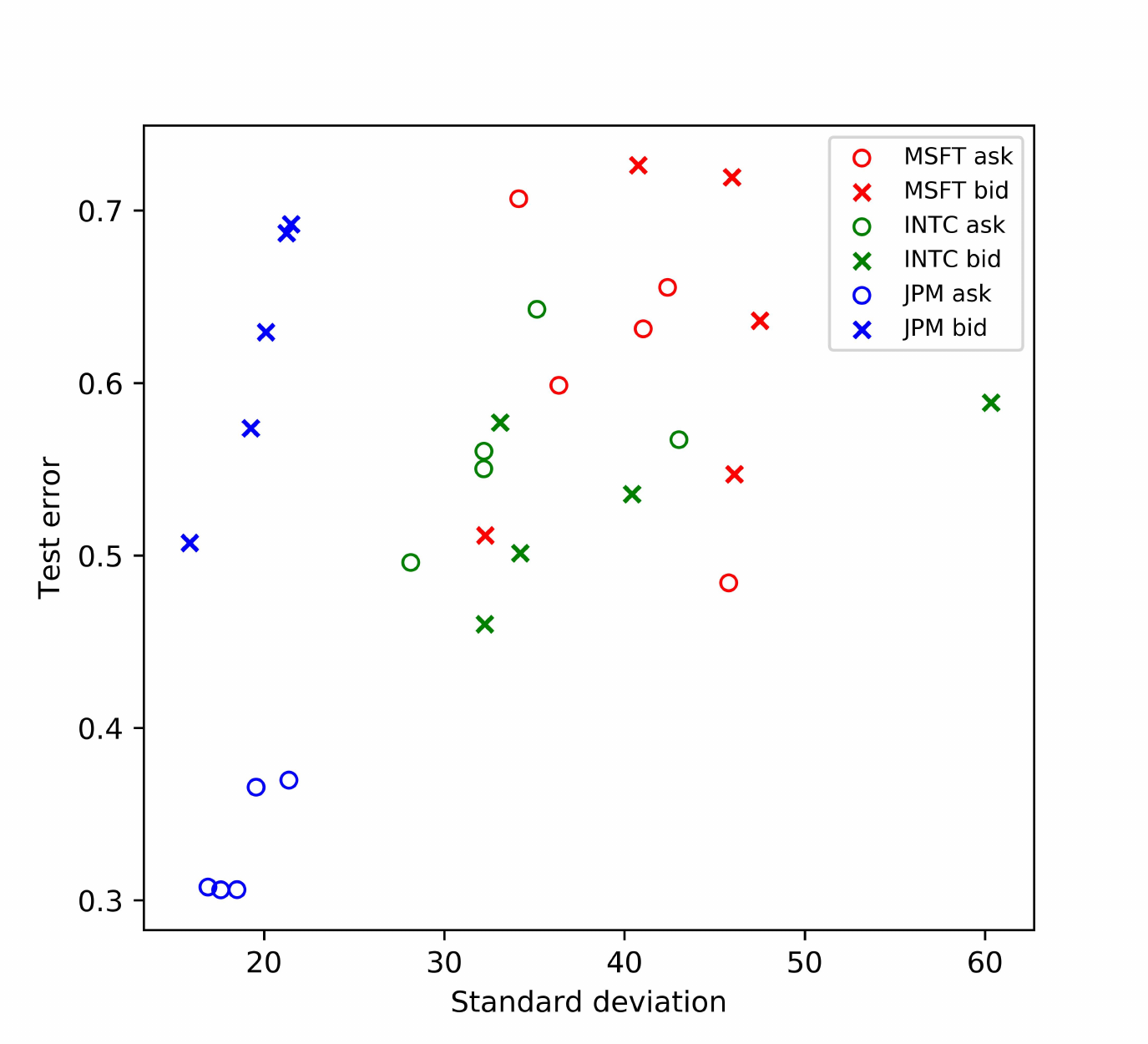}
\vskip -0.5em
\caption{Hourly test loss plotted against hourly volume standard deviation.}
\label{fig:corr}
\end{minipage}
\end{figure}

Table~\ref{tab:continuous} presents evaluation results. Judging from both criteria, we observe that non-linear models outperform linear models (LSVR and RR). The LOBRM family also outperforms traditional non-linear models (especially in terms of R-squared) by effective ensembling of RNNs. 
This indicates that the recurrent structure of RNNs can facilitate the model's capability in explaining temporal variance in LOB volume.
Further, LOBRMs with a  continuous RNN module exhibit superior performance over those with a discrete RNN module in 4 out of 6 experiments. LOBRM (Decay-T) achieves the lowest average L1 test loss and highest average R-squared, indicating that incorporating temporal information in both feature vectors and latent state dynamics is the most suitable approach to capture the model's dependence on time. 
Meanwhile we find the model tends to be overfitting, with validation loss starting to rise before training loss converges. Fig.~\ref{fig:loss} shows loss curves for LOBRM (Decay-T) trained with a higher learning rate of 2e-3. The phenomenon of overfitting also explains why we use a pre-defined exponential decay kernel instead of a fully flexible ODE kernel involving a lot more parameters to model temporal dependence.
Thus, in the following experiments, we continue using LOBRM (Decay-T) as the main model.

As shown in Table~\ref{tab:continuous}, test losses across 6 sets of experiments fluctuate in the range $[0.332,0.628]$ for LOBRM (Decay-T). To test whether this fluctuation results from order volume volatility, we use the pre-trained model to calculate the correlation of (hourly) test L1 loss across all experiments against the (hourly) standard deviation of non-standardized volume resting at deep price levels (see Fig.~\ref{fig:corr}). 
We find that the correlation between hourly L1 loss and volume volatility is significant (p-value\textless0.01) with a correlation coefficient of $\rho=0.48$. Thus, we conclude that LOB prediction accuracy is negatively related to volume volatility.

Compared with the original LOBRM study, which uses a non-chronological training and testing method, we find that the performance of LOBRM trained in a chronological manner is weakened. In order to make the criteria comparable, we first transform the L1 loss in z-score into L1 loss on volume as a percentage of average volume, using $(zscore*std_{i})/mean_{i}$ for each dataset $i$. 
The transformed loss on LOBRM (Decay-T) ranges in $[16.1\%,24.1\%]$ with a mean value of 18.9\%, whereas this criteria in non-chronological experiments has an average of 6.9\%. The main reason lies in that, in non-chronological experiments the model attempts to predict the volume at a target timestep when both volumes at past and future timesteps are known and used for training. 
This leakage of information from the future introduces a form of lookahead bias into the model that tends to increase the prediction accuracy. In the experiments presented here, no future labels are used for training,  thus lookahead bias is eliminated.

We further compare our results of R-squared with results from the statistical model approaching a similar task proposed in \cite{blanchet2017unraveling}, in which presented R-squared values ranged in $[0.81,0.88]$ for daily average volume at different price levels. However, in comparison, our R-squared values presented in Table~\ref{tab:continuous} are regressed against the ground truth with no averaging procedure. As we have only one-day length of data for training, we cannot generate daily average volume prediction for R-squared calculation. When we average volume predictions into hourly frequency, the value of R-squared across 6 experiments ranges in $[0.48,0.88]$ with an average value of 0.71. We find our results comparable to the existing literature, even though those results are not measured using identical conditions.

\subsection{Ablation Study}
\noindent
We conduct an ablation study to demonstrate the effectiveness of module ensembling. Four experiments are conducted: HC, using only the history compiler; ES, using only the market events simulator; HC+ES, which includes the history compiler and event simulator, using a pre-defined weight to combine outputs; and HC+ES+WS, which is the full LOBRM with adaptive weighting scheme.

\begin{table*}[t]
\small
\centering
\caption{Ablation study: test loss.}\vskip -0.5em
\label{tab:ablation}
\setlength\tabcolsep{2mm}
\begin{tabular}{l cc cc cc c}
\toprule
        & \multicolumn{2}{c}{MSFT}  
        & \multicolumn{2}{c}{INTC}
        & \multicolumn{2}{c}{JPM}
        & \multicolumn{1}{c}{Avg.}\\
        \rule{0pt}{2ex}  
        & \multicolumn{1}{c}{Bid} & \multicolumn{1}{c}{Ask} 
        & \multicolumn{1}{c}{Bid} & \multicolumn{1}{c}{Ask} 
        & \multicolumn{1}{c}{Bid} & \multicolumn{1}{c}{Ask}
        & \multicolumn{1}{c}{   } \\
\midrule
HC        &  6.79   &  7.42  &  5.47  &  6.78  &  6.22  &  3.86  & 6.09\\
ES        &  6.31   &  6.99  &  5.94  &  5.70  &  6.04  &  3.36  & 5.72\\
HC+ES     &  6.26   &  6.45  &  5.40  &  5.50  &  6.02  &  3.32  & 5.49\\
HC+ES+WS  &  6.28   &  6.15  &  5.33  &  5.64  &  6.18  &  3.38  & 5.49\\
\bottomrule
\end{tabular}
\end{table*}

Results are shown in Table~\ref{tab:ablation}.
We see that predictions from the HC alone have the highest test error, as the inputs it receives are trimmed to contain only the most relevant data for current LOB volume prediction. 
ES module, by receiving complete TAQ data input and modelling market events as a stochastic process using a continuous RNN, achieves a lower error than HC module and is the dominant module that contributes to prediction accuracy. Combining both HC and ES modules, either using a predefined or adaptive weight, achieves the best performance, which suggets that the predictions from HC and ES are complementary and can be effectively combined to gain a higher accuracy prediction. The purpose of the WS module in the original LOBRM study is to facilitate the model's use in transfer learning and it does not contribute to prediction accuracy when tested on the same stock that it was trained, as is the case here. 

\subsection{Superiority of Sparse Encoding for TAQ}\label{sec:encoding}
\subsubsection{LOB volume prediction} In sparse encoding, only volume information in TAQ is encoded explicitly and price information are embedded implicitly by positions of non-zero elements in one-hot vectors. In explicit encoding, information including price, volume, and trade direction are encoded directly as non-zero elements in feature vectors. Here we use the ES module in LOBRM (Decay-T) as the main model and tested the model prediction accuracy when these two different encoding methods are used. We don't choose the full model as explicitly encoded input cannot be trimmed so HC module is not used. Results are shown in Table~\ref{tab:encoding1}. We see that the model with sparse encoding achieves lower test loss error in 5 out of 6 experiments. The average test loss for the model with sparse encoding is 17.9\% lower than the model with explicit encoding. 

\begin{table*}[t]
\small
\centering
\caption{LOBRM test loss comparison of explicit and sparse encodings of TAQ data.}\vskip -0.5em
\label{tab:encoding1}
\setlength\tabcolsep{2mm}
\begin{tabular}{l cc cc cc c}
\toprule
        & \multicolumn{2}{c}{MSFT}  
        & \multicolumn{2}{c}{INTC}
        & \multicolumn{2}{c}{JPM}
        & \multicolumn{1}{c}{Avg.}\\
        \rule{0pt}{2ex}  
        & \multicolumn{1}{c}{Bid} & \multicolumn{1}{c}{Ask} 
        & \multicolumn{1}{c}{Bid} & \multicolumn{1}{c}{Ask} 
        & \multicolumn{1}{c}{Bid} & \multicolumn{1}{c}{Ask}
        & \multicolumn{1}{c}{   } \\
\midrule
Explicit       &  6.87   &  6.43  &  6.80  &  9.95  &  7.60  &  4.14 &  6.97\\
Sparse         &  6.31   &  6.99  &  5.94  &  5.70  &  6.04  &  3.36 &  5.72\\
\bottomrule
\end{tabular}
\end{table*}

\subsubsection{Price trend prediction} We compare the sparse encoding method with the convolution method proposed in \cite{zhang2019deeplob} for quotes data in the task of stock price trend prediction. On the basis of explicit encoding, the convolution encoding method applies two convolution layers with filters of size $[1 \times 2]$ and stride $[1 \times 2]$ on quote data. 
This structure first convolutionalize price and volume information at ask and bid sides respectively and then convolutionalize two sides' information together. 
We approach a similar task presented in \cite{zhang2019deeplob} but simplify the model structure to concentrate on the prediction accuracy variation brought by different encoding methods. 
We set length of time series samples as $S=50$. For sparse encoding, we set $k=5$. For convolution encoding, we use 16 $[1 \times 2]$ kernels with stride $[1 \times 2]$ followed by LeakyReLU activation. We standardize features using min-max or z-score standardization. Encoded data are passed to an MLP with ReLU activation. A GRU unit is used to receive iterative inputs and the final latent state is connected with an MLP with Softmax activation to generate a  possibility distribution over three labels $\left \{down, same, up\right \}$. We test the model on MSFT five-day dataset, using first three days for training, the fourth day for validation, and the fifth day for testing. We run a rolling average of five timesteps to alleviate label imbalance \cite{ntakaris2018benchmark}, with 29\%, 40\%, and 31\% for \textit{up}, \textit{same}, and \textit{down}. We train the model with cross entropy loss for 50 iterations and choose the model with highest validation accuracy.

Results are shown in Table~\ref{tab:encoding2}.
We can see that the sparse encoding method with two different standardization methods has superior performance in the task of price trend prediction, compared with convolution encoding either with or without price information. Thus, we draw the conclusion that the sparse encoding method for TAQ data can not only benefit the task of LOB volume prediction, but also other tasks including stock price trend prediction.

\begin{table}[t]
\small
\centering
\caption{Future price prediction: validation/test accuracy.}\vskip -0.5em
\label{tab:encoding2}
\setlength\tabcolsep{2mm}
\begin{tabular}{l cc cc cc}
\toprule
        & \multicolumn{1}{c}{Sparse} 
        & \multicolumn{1}{c}{Convolution}
        & \multicolumn{1}{c}{Convolution}\\
        & \multicolumn{1}{c}{(implicit price)} 
        & \multicolumn{1}{c}{(explicit price)}
        & \multicolumn{1}{c}{(no price)}\\
\midrule
Min-max       &  
\multicolumn{1}{c}{59.8\% / 55.2\%}   &  \multicolumn{1}{c}{55.9\% / 54.4\%}  &  \multicolumn{1}{c}{57.2\% / 53.9\%}  \\
Z-score      &  
\multicolumn{1}{c}{60.2\% / 57.4\%}   &  \multicolumn{1}{c}{57.1\% / 54.0\%}  &  \multicolumn{1}{c}{58.3\% / 55.0\%}  \\
\bottomrule
\end{tabular}
\end{table}

\subsection{Is the Model Well-Trained?}
As financial time series suffer from stochastic drift, in the sense that the distribution of data is unstable and tends to vary temporarily, large amounts of data is needed for model training. For example, the LOB used in \cite{blanchet2017unraveling} and \cite{sirignano2019universal} is of one month's and seventeen month's length. 
Here, for all six sets of experiments (3 stocks $\times$ 2 sides) we use three days' LOB data for training, one day's data for validation, and one day's data for testing. We would like to test whether this amount of training data is abundant enough for out-of-sample testing.

\begin{table*}[t]
\small
\centering
\caption{LOBRM (Decay-T) test loss against training size.}\vskip -0.5em
\label{tab:historical}
\setlength\tabcolsep{2mm}
\begin{tabular}{l cc cc cc c}
\toprule
        & \multicolumn{2}{c}{MSFT}  
        & \multicolumn{2}{c}{INTC}
        & \multicolumn{2}{c}{JPM}
        & \multicolumn{1}{c}{Avg.}\\
        \rule{0pt}{2ex}  
        & \multicolumn{1}{c}{Bid} & \multicolumn{1}{c}{Ask} 
        & \multicolumn{1}{c}{Bid} & \multicolumn{1}{c}{Ask} 
        & \multicolumn{1}{c}{Bid} & \multicolumn{1}{c}{Ask}
        & \multicolumn{1}{c}{   } \\
\midrule
Day3       &  7.15   &  7.15  &  6.05  &  6.18  &  6.29  &  3.33  & 6.03\\
Day2+3     &  6.25   &  6.58  &  5.86  &  5.73  &  6.40  &  3.32  & 5.69\\
Day1+2+3   &  6.28   &  6.15  &  5.33  &  5.64  &  6.18  &  3.32  & 5.48\\
\bottomrule
\end{tabular}
\end{table*}

As the dataset we possess contains five consecutive trading days' LOB data, we leave out the fourth day's and fifth day's LOB data for validation and testing. We first use the third day's data for training and then iteratively add in the second and the first day's historical data to observe how the validation and testing loss change. Results are shown in Table~\ref{tab:historical}. We can see that there is a downward tendency in average test loss as more historical data is used for training. This is especially true for {\tt MSFT} ask, {\tt INTC} bid, and {\tt INTC} ask. This phenomenon suggests that the influence of stochastic drift may be alleviated by exposing the model to more historical samples.
Therefore, we would expect even lower out-of-sample errors if more historical data can be used for training.

\section{Conclusion}\label{sec5}
We have extended the research on the LOBRM, the first deep learning model for generating synthetic LOB data. Two major revisions were proposed: standardizing LOB data with time-weighted z-score to improve the model's generalization ability; and substituting the original ODE kernel with an exponential decay kernel (Decay-T) to improve time efficiency. Experiments were conducted on an extended LOBSTER dataset, a strict superset of the data used in the original study, with size approximately ten times larger. Using a fully chronological training and testing regime, we demonstrated that LOBRM (Decay-T) has superior performance over traditional models, and showed the efficacy of module ensembling. We further found that: (1) LOB volume prediction accuracy is negatively related to volume volatility; (2) sparse one-hot positional encoding of TAQ data  can benefit manifold tasks; and (3) there is some evidence that the influence of stochastic drift can be alleviated by increasing the number of historical samples. As a whole, this study validates the use of LOBRM in demanding application scenarios that require efficient inference and involve large amounts of data for training and predicting.

\subsubsection*{Acknowledgements}\label{ack}
Zijian Shi's PhD is supported by a China Scholarship Council (CSC)/University of Bristol joint-funded scholarship. John Cartlidge is sponsored by Refinitiv.

%
%
%
\bibliographystyle{splncs04}
\bibliography{ref.bib}

\end{document}